\documentclass[aps,preprint,showpacs,groupedaddress]{revtex4-1}
\usepackage{amssymb}
\usepackage{mathrsfs}
\usepackage{axodraw4j}
\usepackage{pstricks}
\usepackage{color}
\usepackage{feynMF}
\usepackage{graphicx}
\usepackage{axodraw}

\begin{document}


\title{\bf Robustness of the O($N$) universality class}


\author{William C. Vieira}
\email{william.vieira19@gmail.com}
\affiliation{\it Departamento de F\'\i sica, Universidade Federal do Piau\'\i, 64049-550, Teresina, PI, Brazil}
\author{Paulo R. S. Carvalho}
\email{prscarvalho@ufpi.edu.br}
\affiliation{\it Departamento de F\'\i sica, Universidade Federal do Piau\'\i, 64049-550, Teresina, PI, Brazil}




\begin{abstract}
We calculate the critical exponents for Lorentz-violating O($N$) $\lambda\phi^{4}$ scalar field theories by using two independent methods. In the first situation we renormalize a massless theory by utilizing normalization conditions. An identical task is fulfilled in the second case in a massive version of the same theory, previously renormalized in the BPHZ method in four dimensions. We show that although the renormalization constants, the $\beta$ and anomalous dimensions acquire Lorentz-violating quantum corrections, the outcome for the critical exponents in both methods are identical and furthermore they are equal to their Lorentz-invariant counterparts. Finally we generalize the last two results for all loop levels and we provide symmetry arguments for justifying the latter.
\end{abstract}


\maketitle


\section{Introduction} 

\par The concept of universality class is one of the most solid ideas in the theory of phase transitions and critical phenomena. This idea is a formidably and very well experimentally established fact: that completely distinct physical systems as a fluid and a ferromagnet present the same critical behavior and thus share an identical set of critical exponents \cite{PhysRevLett.28.240, PhysRevLett.28.548, Wilson197475}. Two or more systems are characterized by equal critical indices or equivalently belong to the same universality class when they have in common: their dimension $d$, $N$ and symmetry of some $N$-component order parameter (magnetization for magnetic systems) and if the interactions among their freedom degrees are of short- or long-range type. The O($N$) universality class investigated here generalizes the specific models with short-range interactions: Ising ($N=1$), XY ($N=2$), Heisenberg ($N=3$), self-avoiding random walk ($N=0$), spherical ($N \rightarrow \infty$) etc \cite{Pelissetto2002549}. The influence of the very intuitive parameters $d$ \cite{PhysRevB.86.155112, PhysRevE.71.046112} and $N$ \cite{PhysRevLett.110.141601, Butti2005527, PhysRevB.54.7177} on the critical behavior of many systems was studied. Less intuitive is the effect of the symmetry of the order parameter on the outcome for the critical indices of a given system \cite{PhysRevE.78.061124, Trugenberger2005509}. Investigating the latter subject is the intent of this Letter.

\par The aim of this Letter is to use field-theoretic renormalization group and $\epsilon$-expansion tools for computing the critical exponents for both massless and massive Lorentz-violating (LV) O($N$) self-interacting $\lambda\phi^{4}$ scalar field theories. Nowadays, in the high energy physics scenario, the importance of a LV scalar field theory rests on the possibility of it as a candidate for describing the Higgs sector of the so called LV Standard Model Extension \cite{PhysRevD.84.065030, Carvalho2013850, Carvalho2014320}. On the other hand, in the condensed matter physics realm, the critical behavior of a system is a result of the divergent properties, in a standard Lorentz-invariant (LI) scalar theory, of the Landau-Ginzburg Lagrangian density for some $N$-component self-interacting scalar quantum field. This field is a fluctuating one such that its mean value is proportional to the order parameter of the system. The Lagrangian density necessary for describing the critical properties of the system is composed of just LI and O($N$) relevant operators, i.e. operators invariant under the symmetries above whose canonical dimensions, determined by power counting analysis, are less than or equal to four. High composite operators, with canonical dimensions greater than four, could be considered but they give negligible corrections to the critical behavior. These negligible corrections are known as corrections to irrelevant operators \cite{BrezinZinnJustin9483, BrezinZinnJustin10849, PhysRevD.7.2927}. The initially primitively divergent one-particle irreducible ($1$PI) vertex parts with amputated external legs are then renormalized in a given renormalization scheme. In a general formalism, universality is expressed by the assertion that the critical exponents are universal quantities, i.e. they are identical if they are obtained in a massless or massive theory renormalized in arbitrary or fixed external momenta although these theories be quite different (different renormalization constants, $\beta$ and anomalous dimensions, fixed points etc) in intermediate steps. It can be shown that the mass of the field is connected to the difference of an arbitrary and the critical temperatures. Thus massless and massive theories represent critical and non-critical systems, respectively. In a given renormalized theory we can calculate directly the critical exponents $\eta$ and $\nu$ which are related to the anomalous dimensions of the field $\phi$ and composite field $\phi^{2}$ evaluated at the fixed point, respectively, from the scaling properties of the $1$PI renormalized vertex parts. They are two of the six exponents needed for a complete description of the system. The remaining four exponents can be calculated by using four scaling relations among them. In an approximation where the fluctuations of the field are not taken into account, we have the Landau or mean field approximation for the critical exponents. In this case we have to consider a critical or Gaussian theory in $d\geq 4$ without considering fluctuations. The dimension $4$ is called the critical dimension, the dimension in which and above the theory is superrenormalizable and is not plagued by divergences. In dimensions less than four, a high precision computation of the exponents is needed. This high precision is reached by evaluating their quantum corrections which are beyond their classical Landau values in a divergent but renormalizable theory in $2<d<4$. They are resulting from the fluctuations of the field. In the field theory framework, these corrections correspond to the loop quantum radiative corrections to the $\beta$ and anomalous dimensions of the theory. See the refs. \cite{Kleinert1993545, Kleinert} for a five-loop computation. A LV theory emerges when to the LI standard theory it is included the only LV O($N$) relevant operator \cite{PhysRevD.84.065030}, with canonical dimension equal to four, of the form $K_{\mu\nu}\partial^{\mu}\phi\partial^{\nu}\phi$, where the constant coefficients $K_{\mu\nu}$ (physically playing a role of a constant background field) are dimensionless, symmetric ($K_{\mu\nu} = K_{\nu\mu}$) and equal for all $N$ components of the field such that the O($N$) symmetry of the $N$-component field is preserved. The Lorentz symmetry is slightly violated if these coefficients do not transform as a second order tensor under Lorentz transformations, i.e. $K_{\mu\nu}^{\prime} \neq \Lambda_{\mu}^{\rho}\Lambda_{\nu}^{\sigma} K_{\rho\sigma}$ and for all them we have $|K_{\mu\nu}|\ll 1$. Higher derivative terms could be included but they would give negligible corrections as irrelevant operators. Similar works involving more specific coefficients composed of just their traceless part were published \cite{PhysRevD.7.2911}. Thus the ones considered in this Letter are the most general encompassing all components. In addition, they permit us addressing, for the first time in the literature to our knowledge, the connection between searching a possible modification of the O($N$) universality class by the inclusion of a LV operator and its effect as viewed as a clear violation of the referred symmetry.              

\par In this Letter we obtain explicitly up to next-to-leading order and for all-loop level in
a proof by induction the quantum corrections to critical exponents for both massless and massive O($N$) self-interacting $\lambda\phi^{4}$ scalar field theories with Lorentz violation. For this purpose, we renormalize the massless theory in dimensions less than four by using normalization conditions where the Feynman diagrams are evaluated at fixed nonvanishing external momenta. This fact is a consequence of defining normalization conditions for the $1$PI vertex parts. In this renormalization process only four diagrams are necessary for the desired goal. We perform an identical task by utilizing a massive version of the same theory previously renormalized in four-dimensional space by the counterterm method in minimal subtraction scheme at arbitrary external momenta, namely Bogoliubov-Parasyuk-Hepp-Zimmermann (BPHZ) method \cite{BogoliubovParasyuk, Hepp, Zimmermann}. In this case, for obtaining the critical exponents in the same loop level it was necessary the computation of more than fifteen diagrams and counterterms as will be shown. So the first method is much more simplified. At the end, we generalize the calculation of critical indices for any loop level in both methods and present our conclusions.

\section{Normalization conditions for massless theory: Critical exponents}

\par For determining the quantum corrections to the critical exponents it is necessary considering the divergent properties of a Lagrangian density which describes the system. The Lagrangian density studied here is that of a LV massless O($N$) self-interacting $\lambda\phi^{4}$ scalar field theory
\begin{eqnarray}\label{huytrji}
\mathcal{L}_{0} = \frac{1}{2}(g_{\mu\nu} + K_{\mu\nu})\partial^{\mu}\phi_{0}\partial^{\nu}\phi_{0} + \frac{\lambda_{0}}{4!}\phi_{0}^{4} + \frac{1}{2}t_{0}\phi_{0}^{2}
\end{eqnarray} 
where $\phi_{0}$ and $\lambda_{0}$ are the bare $N$-dimensional vector field in a $d$-dimensional Euclidean space and the bare coupling constant of the self-interacting field, respectively. The last term in the eq. (\ref{huytrji}) is responsible for the generation of composite field $1$PI vertex parts. In Eq.(\ref{huytrji}), all the signs are positive because it is a common sense in Statistical Field Theory to use the theory in the Euclidean spacetime, for the exponentiation of the Lagrangian density looks like the Boltzmann factor of the system in the canonical ensemble formalism, see the reference \cite{Amit}. In Minkowski spacetime, we would have different signs. The critical exponent $\nu$ is furnished by the renormalization of a like mass-term. This is not a problem in a massive theory. But in a massles one, we have to introduce such a composite operator $\phi^{2}\equiv\phi(y)\phi(y)$ evaluated at the same point of spacetime generating perturbatively in terms of $t_{0}$ the bare composite functions $\Gamma_{B}^{(2,1)}$ to be renormalized. These two terms are different but permit us attain the same objective, i.e. computing $\nu$. The notation and conventions used here in the computation of Feynman diagrams are that of the ref. \cite{Amit} based on the original work \cite{BrezinLeGuillouZinnJustin, ZinnJustin}.

\par In a conventional massless $\lambda\phi^{4}$ scalar theory, the renormalized theory is attained if we renormalize the  three primitively divergent vertex functions, namely the two-point $\Gamma^{(2)}$, four-point $\Gamma^{(4)}$ and composite field $\Gamma^{(2,1)}$ functions. Their unrenormalized expansions, with amputated external legs, up to next-to-leading order are   

\begin{eqnarray}\label{gtfrdrdes}
\Gamma^{(2)}_{B} = \quad \parbox{12mm}{\includegraphics[scale=1.0]{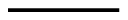}}^{-1} \quad - \quad \frac{1}{6}\hspace{1mm}\parbox{12mm}{\includegraphics[scale=1.0]{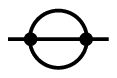}} \quad + \quad \frac{1}{4}\hspace{1mm}\parbox{10mm}{\includegraphics[scale=0.8]{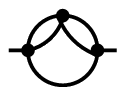}} ,
\end{eqnarray}
\begin{eqnarray}
&&\Gamma^{(4)}_{B} = \quad \parbox{12mm}{\includegraphics[scale=0.09]{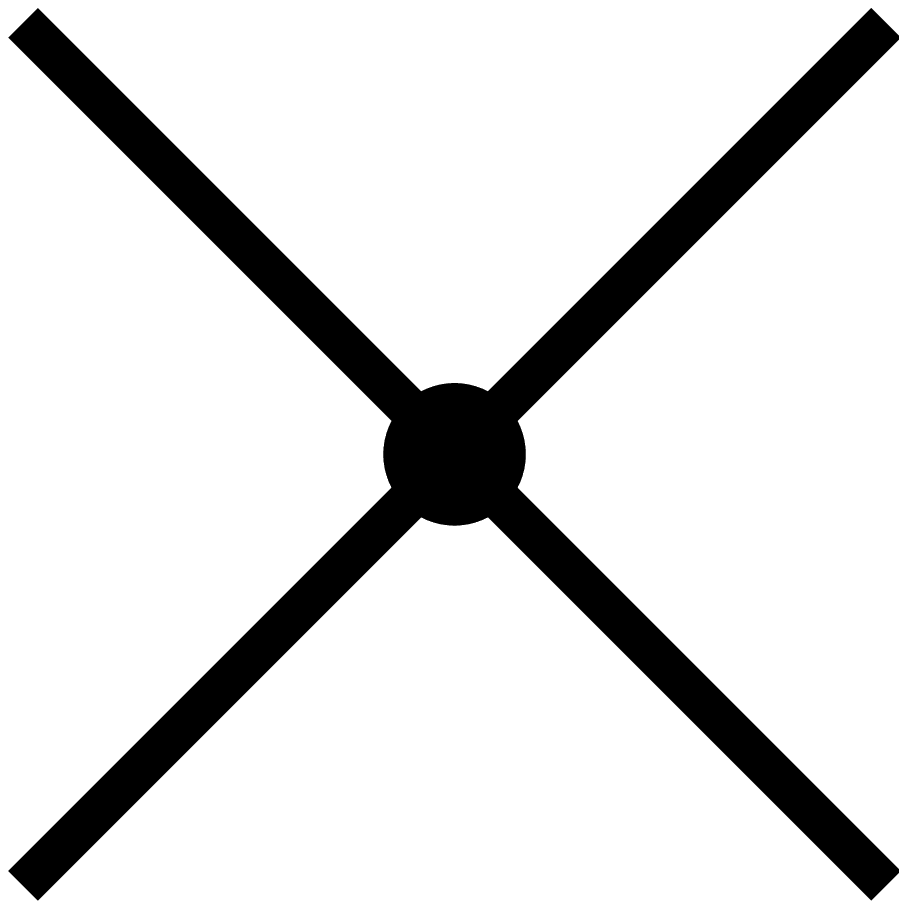}} + \quad \frac{1}{2}\hspace{1mm}\parbox{10mm}{\includegraphics[scale=1.0]{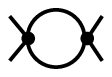}} \quad\quad + 2 \hspace{1mm} perm. \quad + \quad\frac{1}{4}\hspace{1mm}\parbox{16mm}{\includegraphics[scale=1.0]{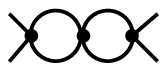}} \quad\quad + 2 \hspace{1mm} perm. \quad + \nonumber \\ && \frac{1}{2}\hspace{1mm}\parbox{12mm}{\includegraphics[scale=0.8]{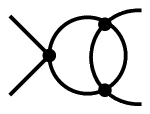}} \quad\quad + 5 \hspace{1mm} perm.,
\end{eqnarray}
\begin{eqnarray}\label{gtfrdesuuji}
\Gamma^{(2,1)}_{B} = \quad \parbox{14mm}{\includegraphics[scale=1.0]{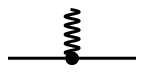}} \quad + \quad \frac{1}{2}\hspace{1mm}\parbox{14mm}{\includegraphics[scale=1.0]{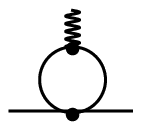}} \quad + \quad \frac{1}{4}\hspace{1mm}\parbox{12mm}{\includegraphics[scale=1.0]{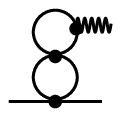}} \quad + \quad \frac{1}{2}\hspace{1mm}\parbox{12mm}{\includegraphics[scale=0.8]{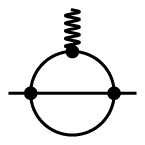}} 
\end{eqnarray}
where \textit{perm.} means a permutation of the external momenta attached to the cut external lines. An internal line with momentum $q$ is the inverse of the free Green function $G_{0}^{-1}(q) = \parbox{12mm}{\includegraphics[scale=1.0]{fig9.eps}}^{-1} = q^{2} + K_{\mu\nu}q^{\mu}q^{\nu}$. As we will see explicitly, the vertex functions above are divergent in $d=4$ and they will be regularized in $d = 4 - \epsilon$. We renormalize them multiplicatively in a specific scheme where the external momenta of the respective renormalized functions are fixed at an arbitrary nonvanishing momentum scale value $\kappa$ (symmetry point (SP)) which renders the Feynman diagrams evaluation easier. This nonzero momentum scale has been chosen because massless diagrams have infrared divergences as powers of the external momenta. The value chosen is such that for an external momentum $P_{i}$, we have $P_{i}\cdot P_{j} = (\kappa^{2}/4)(4\delta_{ij}-1)$ which implies $(P_{i} + P_{j})^{2} \equiv P^{2} = \kappa^{2}$ for $i\neq j$. The variation of the momentum scale or equivalently the renormalization group flow from the bare to the renormalized theory is generated by the correlation length $\xi$. The normalization conditions are
\begin{eqnarray}\label{ygfdxzsze}
\Gamma_{R}^{(2)}(0, g) = 0,
\end{eqnarray}
\begin{eqnarray}
\frac{\partial}{\partial P^{2}}\Gamma_{R}^{(2)}(P, g)\Biggr|_{P^{2} = \kappa^{2}} = 1,
\end{eqnarray}
\begin{eqnarray}
\Gamma_{R}^{(4)}(P, g)|_{SP} = g
\end{eqnarray}
\begin{eqnarray}\label{jijhygtfrd}
\Gamma_{R}^{(2,1)}(P_{1}, P_{2}, Q_{3}, g)|_{SP} = 1
\end{eqnarray}
where $g$ is the renormalized coupling constant and $Q_{3}^{2} = (P_{1} + P_{2})^{2} = \kappa^{2}$ in the eq. (\ref{jijhygtfrd}). A multiplicative renormalization connects the bare and finite $1$PI vertex parts through the field $Z_{\phi}$ and composite field $Z_{\phi^{2}}$ renormalization constants 
\begin{eqnarray}\label{uhygtfrd}
\Gamma_{R}^{(n, l)}(P_{i}, Q_{j}, g, \kappa) = Z_{\phi}^{n/2}Z_{\phi^{2}}^{l}\Gamma_{B}^{(n, l)}(P_{i}, Q_{j}, \lambda_{0}),
\end{eqnarray}
for $(n, l)\neq(0, 2)$ (the function $\Gamma_{B}^{(0, 2)}$, not used here, is renormalized additively), $i = 1, \cdots, n$, $j = 1, \cdots, l$. Defining the dimensionless bare $u_{0}$ and renormalized $u$ coupling constants as $\lambda_{0} = u_{0}\kappa^{\epsilon}$ and $g = u\kappa^{\epsilon}$, we can write the renormalization constants $Z_{\phi}$ and $\overline{Z}_{\phi^{2}} \equiv Z_{\phi}Z_{\phi^{2}}$ perturbatively in powers of $u$.

\par As the bare vertex function is independent of momentum scale $\kappa$, we have $\kappa\frac{\partial}{\partial\kappa}\Gamma_{B}^{(n, l)}\Big|_{\lambda_{0}} = 0$. Applying this relation to the eq. (\ref{uhygtfrd}) we get the renormalization group equation
\begin{eqnarray}
\left( \kappa\frac{\partial}{\partial\kappa} + \beta\frac{\partial}{\partial u} - \frac{1}{2}n\gamma_{\phi} + l\gamma_{\phi^{2}} \right)\Gamma_{R}^{(n, l)} = 0\quad\quad
\end{eqnarray}
where 
\begin{eqnarray}\label{kjjffxdzs}
\beta(u) = \kappa\frac{\partial u}{\partial \kappa} = -\epsilon\left(\frac{\partial\ln u_{0}}{\partial u}\right)^{-1},
\end{eqnarray}
\begin{eqnarray}\label{koiuhygtf}
\gamma_{\phi}(u) = \beta(u)\frac{\partial\ln Z_{\phi}}{\partial u}, \quad \gamma_{\phi^{2}}(u) = -\beta(u)\frac{\partial\ln Z_{\phi^{2}}}{\partial u}.\quad
\end{eqnarray}
It will be more convenient to use the function
\begin{eqnarray}\label{udgygeykoiuhygtf}
\overline{\gamma}_{\phi^{2}}(u) = -\beta(u)\frac{\partial\ln \overline{Z}_{\phi^{2}}}{\partial u} \equiv \gamma_{\phi^{2}}(u) - \gamma_{\phi}(u).
\end{eqnarray}

\par The renormalization constants are calculated in terms of the diagrams evaluated at the symmetry point by imposing the normalization conditions (\ref{ygfdxzsze})-(\ref{jijhygtfrd}) \cite{Amit}. As the $1$PI vertex parts have their external legs cut, the internal bubbles in some diagrams are identical. Thus a few diagrams are proportional between each other if we discard their symmetry factors. This is the case for the one- and two-loop diagrams $\parbox{8mm}{\includegraphics[scale=0.6]{fig14.eps}}  \propto \parbox{6mm}{\includegraphics[scale=0.6]{fig10.eps}}$ \hspace{0.5mm}, $\parbox{8mm}{\includegraphics[scale=0.6]{fig16.eps}} \propto \parbox{10mm}{\includegraphics[scale=0.6]{fig11.eps}} \propto \left(\parbox{6mm}{\includegraphics[scale=0.6]{fig10.eps}}\right)^{2}$, $\parbox{9mm}{\includegraphics[scale=0.5]{fig17.eps}} \propto \hspace{0.5mm}\parbox{8mm}{\includegraphics[scale=0.5]{fig21.eps}}$ for which only two are independent. When we add the two- $\parbox{8mm}{\includegraphics[scale=0.7]{fig6.eps}}$ and three-loop $\parbox{8mm}{\includegraphics[scale=0.6]{fig7.eps}}$ diagrams in our considerations we are left with just four independent diagrams to evaluate. If we write all momenta in the diagrams in units of $\kappa$, the new momenta turn out to be dimensionless and the symmetry point turns $P^{2} = \kappa^{2} \rightarrow 1$ meaning that $P^{\mu} = \kappa\widehat{P}^{\mu} \rightarrow\widehat{P}^{\mu}$ where $\widehat{P}^{\mu}$ is dimensionless and unitary. The dependence on $\kappa$ of diagrams can be absorbed on the coupling constant. As in normalization conditions we have to fix the values of the external momenta, instead of $\parbox{7mm}{\includegraphics[scale=0.6]{fig10.eps}}$, $\parbox{8mm}{\includegraphics[scale=0.7]{fig6.eps}}$, $\parbox{8mm}{\includegraphics[scale=0.5]{fig21.eps}}$ and $\parbox{8mm}{\includegraphics[scale=0.6]{fig7.eps}}$ we have to calculate $\parbox{6mm}{\includegraphics[scale=0.6]{fig10.eps}}_{SP} \equiv \parbox{6mm}{\includegraphics[scale=0.6]{fig10.eps}}\vert_{P^{2}=1}$, $\parbox{8mm}{\includegraphics[scale=0.7]{fig6.eps}}^{\prime} \equiv (\partial \parbox{8mm}{\includegraphics[scale=0.7]{fig6.eps}}/\partial P^{2})\vert_{P^{2}=1}$, $\parbox{8mm}{\includegraphics[scale=0.5]{fig21.eps}}_{SP} \equiv \parbox{8mm}{\includegraphics[scale=0.5]{fig21.eps}}\vert_{P^{2}=1}$ and $\parbox{8mm}{\includegraphics[scale=0.6]{fig7.eps}}^{\prime} \equiv (\partial \parbox{8mm}{\includegraphics[scale=0.6]{fig7.eps}}/\partial P^{2})\vert_{P^{2}=1}$. These diagrams are computed in dimensional regularization in $d = 4 - \epsilon$ \cite{'tHooft1972189, Bollini19725669}. The diagrams $\parbox{7mm}{\includegraphics[scale=0.6]{fig10.eps}}$ and $\parbox{8mm}{\includegraphics[scale=0.5]{fig21.eps}}$ are evaluated perturbatively in the small parameters $K_{\mu\nu}$ up to $\mathcal{O}(K^{2})$ and $\parbox{8mm}{\includegraphics[scale=0.7]{fig6.eps}}$ and $\parbox{7mm}{\includegraphics[scale=0.6]{fig7.eps}}$ up to $\mathcal{O}(K)$ because their calculation would be very tedious in order $\mathcal{O}(K^{2})$ \cite{Carvalho2014320, Ramond}. Finally, for $N = 1$ and without taking into account their symmetry factors, we have the results
\begin{eqnarray}
\parbox{10mm}{\includegraphics[scale=1.0]{fig10.eps}}_{SP} = \frac{1}{\epsilon}\Biggl[ \left(1 + \frac{1}{2}\epsilon \right)\Pi -\frac{1}{2}\epsilon K_{\mu\nu}\widehat{P}^{\mu}\widehat{P}^{\nu} + \frac{1}{4}\epsilon K_{\mu\nu}K_{\rho\sigma}(\widehat{P}^{\mu}\widehat{P}^{\nu}\delta^{\rho\sigma} + \widehat{P}^{\mu}\widehat{P}^{\nu}\widehat{P}^{\rho}\widehat{P}^{\sigma}) \Biggl],\quad
\end{eqnarray}   
\begin{eqnarray}
\parbox{12mm}{\includegraphics[scale=1.0]{fig6.eps}}^{\prime} = -\frac{1}{8\epsilon}\Biggl[ \left(1 + \frac{5}{4}\epsilon \right)\Pi^{2} -\epsilon K_{\mu\nu}\widehat{P}^{\mu}\widehat{P}^{\nu} \Biggl],
\end{eqnarray}  
\begin{eqnarray}
\parbox{10mm}{\includegraphics[scale=0.9]{fig7.eps}}^{\prime} = -\frac{1}{6\epsilon^{2}}\Biggl[ (1 + 2\epsilon )\Pi^{3} -\frac{3}{2}\epsilon K_{\mu\nu}\widehat{P}^{\mu}\widehat{P}^{\nu} \Biggl],
\end{eqnarray}  
\begin{eqnarray}
&&\parbox{12mm}{\includegraphics[scale=0.8]{fig21.eps}}_{SP} = \frac{1}{2\epsilon^{2}}\Biggl[ \left(1 + \frac{3}{2}\epsilon \right)\Pi^{2} -\epsilon K_{\mu\nu}\widehat{P}^{\mu}\widehat{P}^{\nu} + \frac{1}{2}\epsilon K_{\mu\nu}K_{\rho\sigma}(\widehat{P}^{\mu}\widehat{P}^{\nu}\delta^{\rho\sigma} + \widehat{P}^{\mu}\widehat{P}^{\nu}\widehat{P}^{\rho}\widehat{P}^{\sigma}) \Biggl]\quad
\end{eqnarray}  
where
\begin{eqnarray}
\Pi = 1 - \frac{1}{2}K_{\mu\nu}\delta^{\mu\nu} + \frac{1}{8}K_{\mu\nu}K_{\rho\sigma}\delta^{\{\mu\nu}\delta^{\rho\sigma\}} + ...,
\end{eqnarray}
$\delta^{\{\mu\nu}\delta^{\rho\sigma\}} \equiv \delta^{\mu\nu}\delta^{\rho\sigma} + \delta^{\mu\rho}\delta^{\nu\sigma} + \delta^{\mu\sigma}\delta^{\nu\rho}$ and $\delta^{\mu\nu}$ is the Kronecker delta symbol. In the renormalization process, the poles in $\epsilon$ disappear and the $\beta$ function, field and composite field anomalous dimensions turn out to be finite as required by the renormalization prescription. Also, the dimensionless terms 
\begin{eqnarray}\label{uhdygdyge}
K_{\mu\nu}\widehat{P}^{\mu}\widehat{P}^{\nu}, \quad K_{\mu\nu}K_{\rho\sigma}\widehat{P}^{\mu}\widehat{P}^{\nu}\widehat{P}^{\sigma}\widehat{P}^{\rho}
\end{eqnarray}
are cancelled out and the dependence of these functions on $K$ is only due to $\Pi$. This fact will be essential to our results. Thus they assume the form 


\begin{eqnarray}
\beta(u) = u\left[ -\epsilon + \frac{N + 8}{6}\left( 1 + \frac{1}{2}\epsilon \right)\Pi u - \frac{3N + 14}{12}\Pi^{2}u^{2}\right],
\end{eqnarray}
\begin{eqnarray}
\gamma_{\phi}(u) = \frac{N + 2}{72}\left[ \left( 1 + \frac{5}{4}\epsilon \right)\Pi^{2}u^{2} - \frac{N + 8}{12}\Pi^{3}u^{3} \right],\quad\quad
\end{eqnarray}
\begin{eqnarray}
\overline{\gamma}_{\phi^{2}}(u) = \frac{N + 2}{6}\left[ \left( 1 + \frac{1}{2}\epsilon \right)\Pi u - \frac{1}{2}\Pi^{2}u^{2} \right].
\end{eqnarray}
Observe that they can be obtained through the LI ones if we make the transformation $u^{(0)}=\Pi u$ in the latter ones, where $u^{(0)}$ is the LI dimensionless renormalized coupling constant \cite{Amit, BrezinLeGuillouZinnJustin, ZinnJustin}. This is only possible owing to the cancelling of the dimensionless terms in the eq. (\ref{uhdygdyge}). In fact, it was shown, by a well-known coordinate redefinition, that we can remove the coefficients $K_{\mu\nu}$ from the original LV Lagrangian density and define an effective LI theory with its coupling constant scaled by the $\Pi$ factor \cite{PhysRevD.84.065030}. Thus the effective dimensionless renormalized coupling constant, $\Pi u$, for the LV theory is identified with the corresponding LI coupling constant $u^{(0)}$. The Gaussian and nontrivial fixed points are achieved by the condition $\beta(u^{*}) = 0$. The Gaussian fixed point is the trivial solution $u^{*} = 0$ and gives results for the critical exponents in the mean field approximation. The corrections to Landau theory are given by using the nontrivial fixed point evaluated from the equation in the bracket, namely   
\begin{eqnarray}
u^{*} = \frac{6\epsilon}{(N + 8)\Pi}\left\{ 1 + \epsilon\left[ \frac{3(3N + 14)}{(N + 8)^{2}} -\frac{1}{2} \right]\right\}.\quad\quad
\end{eqnarray}
We see that $u^{*} \equiv u^{*(0)}/\Pi$, as discussed above. Then we find that the critical indices $\eta\equiv\gamma_{\phi}(u^{*})$
and $\nu^{-1}\equiv 2 - \eta - \overline{\gamma}_{\phi^{2}}(u^{*})$
are identical to that of the corresponding LI theory, namely $\eta\equiv\eta^{(0)}$ and $\nu\equiv\nu^{(0)}$ \cite{Wilson197475}. Mathematically, this occurs in virtue of the functional form of the $\beta$, Wilson functions and nontrivial fixed point. They are such that the LV $\Pi$ factor is cancelled out in the critical exponents calculation procedure. Now we present the computation of the same exponents in a massive theory in another renormalization scheme.

\section{Critical exponents in the BPHZ method}

\par In previous works, for a four-dimensional massive version of the theory studied here, the two-loop quantum contributions to the $\beta$ function, field \cite{PhysRevD.84.065030} and mass anomalous dimensions \cite{Carvalho2013850} were calculated in the BPHZ method. A more accurate determination of the field anomalous dimension was given by the evaluation of its three-loop term \cite{Carvalho2014320}. In this method, the $\beta$ function below the critical dimension, known as Gell-Mann-Low function, is composed of the extra term $-\epsilon u$ and its loop quantum corrections \cite{Brezin, Vladimirov, Naud}. After the computation of more than fifteen diagrams and counterterms, the $\beta$ function, field and mass anomalous dimensions up to next-to-leading order read
\begin{eqnarray}
&&\beta(u) = u\left[ -\epsilon + \frac{N + 8}{3(4\pi)^{2}}\Pi u - \frac{3N + 14}{3(4\pi)^{4}}\Pi^{2}u^{2}\right],
\end{eqnarray}
\begin{eqnarray}
\gamma(u) = \frac{N + 2}{36}\left[ \frac{1}{(4\pi)^{4}}\Pi^{2}u^{2} - \frac{N + 8}{12(4\pi)^{6}}\Pi^{3}u^{3} \right],\quad\quad
\end{eqnarray}
\begin{eqnarray}
\gamma_{m}(u) = \frac{N + 2}{6}\left[ \frac{1}{(4\pi)^{2}}\Pi u - \frac{5}{6(4\pi)^{4}}\Pi^{2}u^{2} \right].
\end{eqnarray}
We can also evaluate the nontrivial fixed point, with $\overline{u}\equiv u/(4\pi)^{2}$, and get the result
\begin{eqnarray}
\overline{u}^{*} = \frac{6\epsilon}{(N + 8)\Pi}\left\{ 1 + \epsilon\left[ \frac{3(3N + 14)}{(N + 8)^{2}} \right]\right\}.
\end{eqnarray}
Following the notation in the ref. \cite{Kleinert}, we also obtain the relation $\overline{u}^{*} \equiv \overline{u}^{*(0)}/\Pi$ in this renormalization scheme and get the same LI critical exponents $\eta$ and $\nu$ as earlier. This shows their universal character as being identical in both methods. Now we present an all-loop generalization for the critical exponents in both methods. 

\section{All-loop critical exponents in the normalization conditions and in the BPHZ methods}

\par It was also shown, in four-dimensional space, that the $\beta$ function and the field $\gamma$ and mass $\gamma_{m}$ anomalous dimensions in the BPHZ method (analogous arguments can be used for showing similar results for the $\beta$, $\gamma_{\phi}$ and $\overline{\gamma}_{\phi^{2}}$ functions in the normalization conditions method) can be written for all-loop order \cite{PhysRevD.84.065030, Carvalho2013850, Carvalho2014320}. As discussed, in $d<4$ we have
\begin{eqnarray}\label{uhgufhduhufdhu}
\beta(u) = u\left[ -\epsilon + \sum_{n=2}^{\infty}\beta_{n}^{(0)}\Pi^{n-1}u^{n-1}\right], 
\end{eqnarray}
\begin{eqnarray}
\gamma(u) = \sum_{n=2}^{\infty}\gamma_{n}^{(0)}\Pi^{n}u^{n},
\end{eqnarray}
\begin{eqnarray}
\gamma_{m}(u) = \sum_{n=1}^{\infty}\gamma_{m,n}^{(0)}\Pi^{n}u^{n}.
\end{eqnarray}
where $\beta_{n}^{(0)}$, $\gamma_{n}^{(0)}$ and $\gamma_{m,n}^{(0)}$ are the corresponding nth-loop quantum corrections to the respective functions. One more time we have the Gaussian fixed point as the trivial solution $u^{*}=0$. For the nontrivial fixed point, we get the nontrivial solution $u^{*} = u^{*(0)}/\Pi$ from the eq. (\ref{uhgufhduhufdhu}) in the bracket, where $ u^{*(0)}$ is the LI nontrivial fixed point for all-loop level. The LV $\Pi$ factor cancels out once again and we get the LI critical indices $\eta$ and $\nu$. The power and generality of this result is that the critical exponents are invariant under LV transformations in the spacetime in which the field is defined for any loop level.

\section{Conclusions}

\par We have explicitly computed up to next-to-leading order and obtained for all-loop level in a proof by induction the critical exponents for Lorentz-violating O($N$) $\lambda\phi^{4}$ scalar field theories by using two independent field-theoretic renormalization group methods combined with dimensional regularization and $\epsilon$-expansion techniques. In both methods we have obtained identical critical indices, as required by universality requirements, and furthermore we have found that they are equal to their LI counterparts. The latter result says us that although the renormalization constants, the $\beta$ and anomalous dimensions acquire Lorentz-violating quantum corrections, the outcome for the critical exponents was not affected by the introduction of the LV kinetic term in the Lagrangian density of the system. As it is known, the resulting effect of an operator on the critical behavior of a system or equivalently its relevance is not previously ensured by naive power-counting, but otherwise it has to be checked after the renormalization process \cite{Amit}. Physically, this result is understood if we realize that the broken symmetry here is not an internal symmetry of the field (which would change the universality class of the system), but a spacetime symmetry in which the field is embedded \cite{Aharony}. So the critical exponents are insensible to this symmetry breaking mechanism. This result leaves our confidence greater than ever on the robustness of the statement ensuring that a set of distinct physical systems belong to the same universality class when they share identical dimension $d$, $N$ and symmetry of their $N$-component order parameters if only short-range interactions are present. Moreover, the present Letter opens new issues on the role played by a theory with a broken spacetime symmetry on the description of universal quantities other than bulk critical exponents such as critical exponents in geometries subjected to different boundary conditions, susceptibility and specific heat amplitude ratios, scaling equations of state etc.

\bibliography{apstemplate}

\end{document}